\documentclass[aps,pra,twocolumn,groupedaddress,showpacs,showkeys]{revtex4}

\usepackage{graphicx}
\begin{document}
%\begin{flushright}
%DSF- 15/2004  \\ quant-ph/yymmnnn\\
%\end{flushright}
\vspace{1cm}
\newcommand{\be}{\begin{equation}}
\newcommand{\ee}{\end{equation}}

\title{Neutron-antineutron transition as a test-bed for dynamical CPT violations}

\author{ Andrea Addazi}
%\thanks{}
\affiliation{Dipartimento di Fisica,
 Universit\`a di L'Aquila, 67010 Coppito AQ and
LNGS, Laboratori Nazionali del Gran Sasso, 67010 Assergi AQ, Italy }

\date{\today}

\begin{abstract}

We show a simple mechanism 
for a dynamical CPT violation in the neutron sector.
In particular, we show a {\it CPT-violating see-saw mechanism}, 
generating a Majorana mass and a CPT violating mass for the neutron.
CPT-violating see-saw involves a 
 sterile partner of the neutron, 
living in a hidden sector,
in which CPT is spontaneously broken. 
In particular, neutrons (antineutrons) can communicate with 
the hidden sector through non-perturbative quantum gravity effects 
called
{\it exotic instantons}. Exotic instantons dynamically 
break R-parity, generating one effective vertex between the neutron 
and its sterile partner. 
In this way, we show how a small CPT violating mass term for the neutron
 is naturally generated.
This model can be tested in the next generation 
of experiments in neutron-antineutron physics. 
This strongly motivates researches of CPT-violating effects in neutron-antineutron physics,
as a test-bed for dynamical CPT-violations in SM.

\end{abstract}

% insert suggested keywords - APS authors don't need to do thisy to

\maketitle

\section{Introduction}
Is CPT a fundamental symmetry of Nature?

Only experimental observables can answer to 
this crucial question. 
CPT is strictly connected with other deep issues 
like Lorentz invariance (LI), Causality and Locality.
For this motivation, CPT seems 
an untouchable symmetry of Quantum Field Theories (QFT).
However, it is commonly retained that QFT
are effective theories of a more fundamental one, 
including quantum gravity. 
In this generic idea, locality or LI could be 
emergent/approximated principles rather than fundamental ones
\footnote{To Break Lorentz Invariance can be problematic also at classical level.
For an interesting explicit example, in \cite{Capozziello} 
geodetic instabilities of Classical Lorentz Breaking Massive Gravity
were discussed. A discussion of these effects 
is often neglected in literature of Modified Gravities and Massive gravities. 
On the other hand, issues about problematic acausal divergences at 
quantum level for Non-local QFTs were discussed in \cite{Esposito} . 
In this last case, $\mathcal{N}=1$ supersymmetry seems to eliminate many acausal diagrams, 
but not other many ones, 
as carefully checked in \cite{Esposito}. Probably, $\mathcal{N}=2$ susy can ulteriorly alleviate acausal divergences.
In \cite{Addazi:2015ppa}, how the formation of classicalons can avoid acausal divergences in 
scattering amplitudes is discussed. 
}. 
On the other hand, CPT could be spontaneously or dynamically
broken even if starting from a CPT-preserving theory. 
For example, one can envisage the presence of a hidden sector, 
in which CPT is spontaneously broken.
In this case, mediators can transmit informations of CPT-violations (CPTV)
from the hidden sector to Standard Model (SM). 
CPTV can manifest
itself in particle-antiparticle mass differences (even if it is not the only CPTV observable, in general).
For general papers and reviews on CPT-violations, 
see \cite{CPTV}.
Kaons (Antikaons) are particularly sensitive to 
CPTV effects, so that the actual limits
are very stringent: $R_{K}=|\Delta m_{K\bar K}|/m_{K}<8 \times 10^{-19}$
\cite{CLEPER}. 
On the other hand, limits on neutron (antineutron) are 
much milder than the previous ones: 
only $R_{n}=|\Delta m_{n\bar{n}}|/m_{n}<(9\pm 5)\times 10^{-5}$
\cite{Baldo}. Next generation of experiments 
will improve limits on neutron-antineutron physics \cite{Next}.
In \cite{AJO}, it was proposed that 
if a neutron-antineutron transition was found, 
it would be a test for CPT. 
See also
 \cite{BM} for a recent paper on CPTV in $n-\bar{n}$ transitions.
 
In this paper, we propose a simple mechanism, 
dynamically breaking CPT in SM sector, 
generating a CPT violating mass term only for the neutron!
This mechanism can avoid stringent limits on kaons as well as for other possible channels.
We propose a {\it CPTV see-saw mechanism for the neutron} \footnote{
The first proposal of a see-saw type I mechanism for the neutron (CPT-preserving)
was shown by Berezhiani {\it et al}
in \cite{Zurab}, for neutron-mirror neutron oscillations.
Clearly, Berezhiani's diagram can be considered 
also for $n-\bar{n}$. In this diagram, 
$\psi$ is substituted by a sort of RH neutron,
with baryon number $B=-1$ and a B-violating Majorana mass. In this model, 
they assume that other possible gauge invariant interactions 
of the RH neutrons with SM particles are not allowed. 
Considerations done in this paper can also be applied for this model.  
This model could be fully justified by appropriate extra discrete symmetries or by exotic instantons in non-susy intersecting D-branes' models. }, 
in which a sterile partner of the neutron, called "neutronic weight" or briefly "nweight", 
living in a "CPTV hidden sector",
 can generate a Majorana mass and a CPT violating mass for the neutron.
In a "CPTV hidden sector",
we assume that CPT is violated by an unspecified 
mechanism, generating an unsuppressed CPTV mass
for the nweight. 
Non-perturbative stringy effects, 
called {\it exotic instantons}, can provide privileged portals from the hidden sector 
to neutrons.
For these motivations, a CPTV in neutron physics seems 
strongly sustained by 
 (exotic) instantons-mediated models.
  \begin{figure}[t]
\centerline{ \includegraphics [height=4cm,width=1\columnwidth]{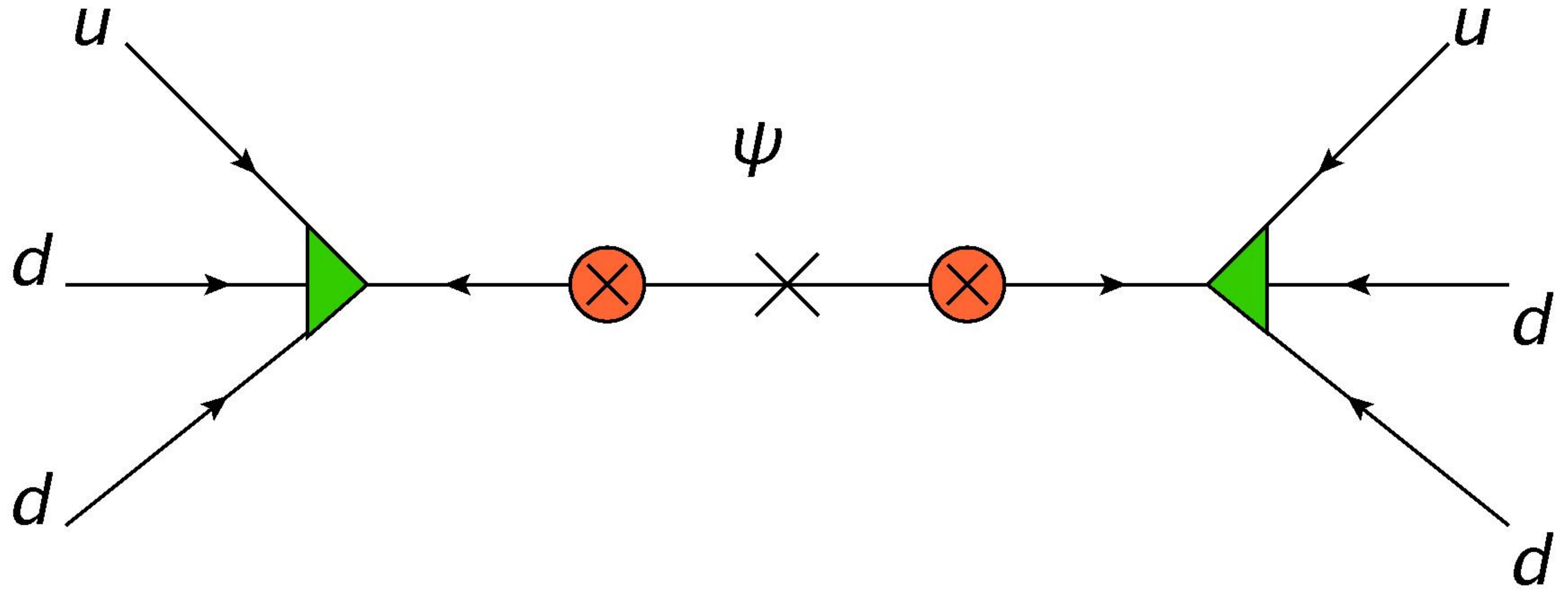}}
\vspace*{-1ex}
\caption{Diagram of a CPT-violating see-saw mechanism for the neutron. 
$u,d,d$ are right-handed up and down quarks, while $\psi$ is a partner of the neutron, named "neutronic weight" or briefly "nweight". 
Black cross is a Majorana mass for $\psi$, while 
red-crosses are CPTV mass terms for $\psi$.
As a consequence, $\psi,\bar{\psi}$ get a CPT violating mass matrix.
As a CPTV see-saw mechanism, also a 
CPT violating mass matrix for $n,\bar{n}$
is generated. The 'green' vertices 
can be directly generated by Exotic Instantons.}
\label{plot}   % \ref{plot}
\end{figure}
\section{CPTV see-saw mechanism for the neutron}
Let us introduce the basic idea of our model.
We will justify our assumptions later. 
We introduce a fermionic field $\psi$
living in a hidden sector. 
We call such a fermion "neutronic weight" 
or "nweight". 
Such a particle is a singlet with respect to
SM gauge group (but not necessary for other hidden gauge groups), {\it i.e} $\psi(1,1,0)$ of $SU(3)\times SU(2) \times U(1)$.
We assume that $\psi$ has a Majorana and a Dirac Mass term, 
$m_{\psi}\bar{\psi}\psi+\mu_{\psi}\psi^{2}+h.c$. 
However, $\psi$ has also a large CPT-violating mass term 
$ \Lambda_{CPTV}\psi^{\dagger}\psi+h.c$.
Such a mass term is associated to a local Lorentz symmetry 
breaking in the hidden sector:
\begin{equation}
\label{LIV}
a_{\mu}\bar{\psi}\gamma^{\mu}\psi \rightarrow  \Lambda_{CPTV}\psi^{\dagger}\psi
\end{equation}
where $a_{\mu}$ is a spurion vector field getting an expectation value 
$\langle a_{0} \rangle$
so that $\langle a_{0}\rangle \bar{\psi}\gamma_{0}\psi=\Lambda_{CPTV}\psi^{\dagger}\psi$.
At this point, one could be 'afraid' about the fact that such a term
seems gauge equivalent to zero such as an electric potential. However, in a multifermionic 
theory for the hidden sector such a conclusion is in general not true,
as discussed in litterature cited above \footnote{
Because of this, signals in neutron physics have to be dependent on the momentum of neutrons (antineutrons).
However, in our paper, we are interested to low energy observables, so that momentum-dependent effects
are completely negligible. On the other hand, in the opposite regime, 
momentum dependent effects in neutron physics would 
afflict neutron(antineutron) propagation.
Unfortunately, it is not simple to try an interesting example in phenomenology in which such a test could be 
realistically done. With the increasing of statistics in UHECR, one could limit such Lorentz/CPT violations in future,
if a consistent part of UHECR was composed of protons.   }-\footnote{In principle, one could introduce other possible Lorentz violating couplings for $\psi$
such as $b_{\mu}\bar{\psi}\gamma^{\mu}\gamma_{5}\psi$ (with $b_{\mu}$ a spurion field getting a LIV expectation value) and so on, as usually proposed in literature mentioned above. These other possible operators
would affect neutron (antineutron) physics through our see-saw mechanism. In this paper, we have focalized 
on the case in which the only relevant LIV OPE is $a_{\mu}\bar{\psi}\gamma^{\mu}\psi$. 
A more general case will deserve future investigations beyond the purposes of this paper. 
}.

Now, let us consider the following effective interactions
\begin{equation}
\label{eff}
{\mathcal{L}}_{eff}={\mathcal{L}}_{1}+{\mathcal{L}}_{2}
\end{equation}
where
\begin{equation}
\label{two}
{\mathcal{L}}_{1}=\mu_{n\psi}n\psi+h.c
\end{equation}
\begin{equation}
\label{three}
{\mathcal{L}}_{2}=m_{\psi}\bar{\psi}\psi+\mu_{\psi} \psi \psi+\Lambda_{CPTV}\psi^{\dagger}\psi+h.c
\end{equation}
${\mathcal{L}}_{1}$ is an effective vertex of the neutron and the nweight field, 
as a portal to the hidden sector. We will discuss later 
how can be possible the generation of such an effective 
vertex, without generating other dangerous vertices. 
Lagrangian 
${\mathcal{L}}_{eff}$ can give rise to a CPT violating mass term $\mu_{CPTV}n^{\dagger}n+h.c$.
This is a simple example of a CPT-violating see-saw mechanism 
for the neutron. The associated diagram is shown in Fig.1.
In particular, $\psi,\bar{\psi}$ have a non-diagonal mass matrix 
\begin{equation}
\label{MCPT}
\mathcal{M}_{\psi\bar{\psi}} = \left( \begin{array}{cc} m_{\psi}+\Lambda_{CPTV} 
 & \mu_{\psi}
\ \\ \mu_{\psi}^{*} & m_{\psi}-\Lambda_{CPTV} \ \\
\end{array} \right) \end{equation}
On the other hand, this is a sub-matrix of 
neutron-nweight system.
In basis $(n,\bar{n},\psi,\bar{\psi})$,
the complete system is 
\begin{equation}
\label{MCPT}
\mathcal{M}_{CPTV} = \left( \begin{array}{cc} 
\mathcal{M}_{n\bar{n}}^{0} & \mathcal{M}_{n\psi}
 \ \\
 \mathcal{M}_{n\psi}^{\dagger} & \mathcal{M}_{\psi\bar{\psi}}
\end{array} \right) \end{equation}
with 
\begin{equation}
\label{MCPT2}
\mathcal{M}_{n\psi} = \left( \begin{array}{cc} 
\mu_{n\psi} & 0
 \ \\ 
0 & \mu_{n\psi}^{*}
\end{array} \right) \end{equation}
and $\mathcal{M}^{0}_{n\bar{n}}=m_{n}\mathcal{I}_{2\times 2}$. 

The general CPTV see-saw formula for the neutron
is
\begin{equation}
\label{Formula}
\mathcal{M}_{n\bar{n}}=-\mathcal{M}_{n\psi}^{\dagger}\mathcal{M}_{\psi\bar{\psi}}^{-1}\mathcal{M}_{n\psi}
\end{equation}
Let us note that if $\Lambda_{CPTV}=m_{\psi}=0$, we recover 
the CPT preserving case with a Majorana nweight.
In this case, $|m_{n}-m_{\bar{n}}|=0$, but a Majorana mass
for the neutron (antineutron) is generated.
Assuming $\mu_{\psi}>>m_{n}$, $\mu_{n\bar{n}}\simeq \mu_{n\psi}^{2}/\mu_{\psi}$. 
On the other hand, if nweight has also a Dirac mass term,
but again in CPT-preserving phase $\Lambda_{CPTV}=0$, 
we obtain a Majorana mass 
$\mu_{n\bar{n}}=-\mu_{n\psi}^{2}\mu_{\psi}/(m_{\psi}^{2}-\mu_{\psi}^{2})$.
Now, let us consider CPTV case $\Lambda_{CPTV}\neq 0$.
In this case, not only the neutron Majorana mass 
is non-trivially affected,
but we also obtain a difference in 
neutron and antineutron Dirac masses:
\begin{equation}
\label{e1}
\mu_{n\bar{n}}=-\frac{\mu_{n\psi}^{2}\mu_{\psi}}{-\Lambda_{CPTV}^{2}+m_{\psi}^{2}-\mu_{\psi}^{2}}
\end{equation}
\begin{equation}
\label{e2}
|m_{n}-m_{\bar{n}}|=\frac{2\mu_{n\psi}^{2}\Lambda_{CPTV}}{|\Lambda_{CPTV}^{2}-m_{\psi}^{2}+\mu_{\psi}^{2}|}
\end{equation}

\section{Theoretical interpretations of the CPTV-seesaw model. }

 \begin{figure}[t]
\centerline{ \includegraphics [height=3.0cm,width=1.0\columnwidth]{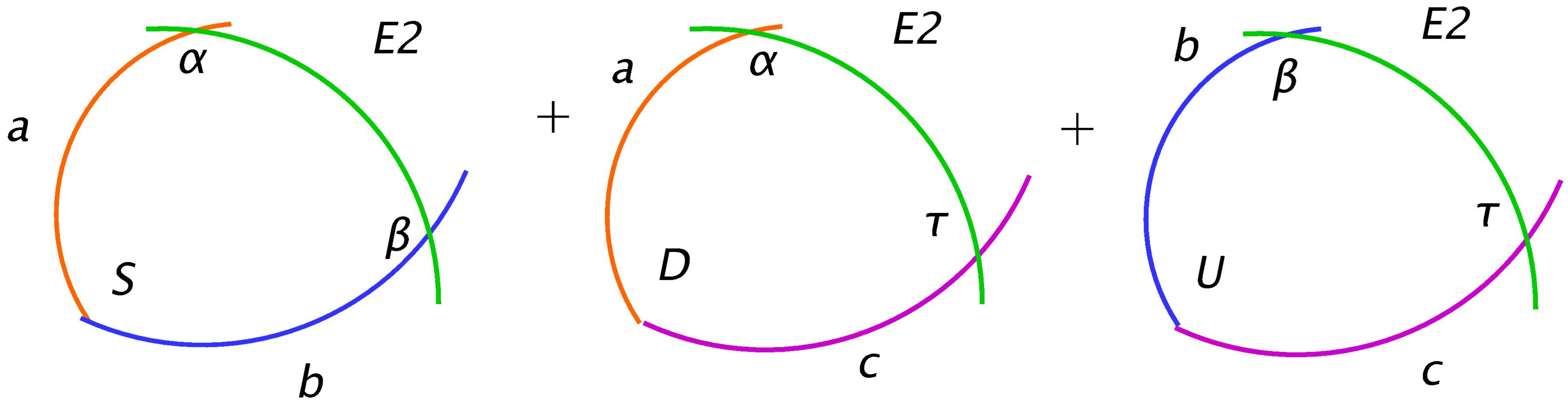}}
\vspace*{-1ex}
\caption{ Mixed disk amplitudes generating an effective $n-\psi$-vertex. The $E2$-instanton
is in green, and it intersects ordinary D6-branes' stacks. This generates the desired effective interactions 
among ordinary fields and modulini.   }
\label{plot}   % \ref{plot}
\end{figure}

The main theoretical problems behind 
our toy-model are: why 
does such a singlet field interact only with 
neutrons? 
Can such a field destabilize nuclei?
Can CPTV phases be transmitted also in kaons or other well constrained channels?

In this section, we will comment how possible can be a situation in which 
$\psi$ has one and only one portal-operator 
like $\mathcal{O}_{n\psi}=\psi u^{c}d^{c}d^{c}/\Lambda_{n\psi}^{2}$,
without generating a plethora of other dangerous operators. 
In other words, we desire a mechanism to dynamically break baryon number 
without generating all possible baryon and lepton violating operators.
As shown in
 \cite{E1,E2},
 $(B-L)$-parity  can be {\it dynamically} broken from
 non-perturbative quantum gravity effects known as  
 {\it exotic stringy instantons}.
 This class of instantons corresponds to 
 Euclidean D-branes (or E-branes), 
 intersecting physical D-brane stacks.
 More precisely, in IIA string-theory \footnote{Let
 us remind some useful seminal papers on open string theories in
\cite{Sagnotti}.}
 , the class of "exotic instantons" 
 corresponds to $E2$-branes wrapping different 
 3-cycles on the Calabi-Yau compactification with respect to ordinary $D6$-branes
\footnote{However, there are other classes of instantons that may be relevant for our purposes.
These classes were studied in \cite{Parsa}. I would like to thank Parsa Ghorbani for discussions on these aspects. }. 

In intersecting D-branes' models, Standard Model content and 
the extra singlet can be easily reconstruct in the low energy limit. 
On the other hand, an $E2$-instanton, with the appropriate 
intersections and Chan-Paton group, can generate a superpotential 
like $\mathcal{W}_{n\psi}=S U^{c}D^{c}D^{c}/\mathcal{M}_{0}$
where $\mathcal{M}_{0}=e^{+S_{E2}}M_{S}$, where $M_{S}$ is the string scale
and $e^{-S_{E2}}$ is a function of geometric moduli associated to 
3-cycles of $E2$-brane on $CY_{3}$. 
Now, $S$ is a nweight superfield with $\psi$ as the lowest component.
These superpotential terms generate 
operator $\mathcal{O}_{n\psi}=\psi u^{c}d^{c}d^{c}/\Lambda_{n\psi}^{2}$
with $\Lambda_{n\psi}^{2}=\mathcal{M}_{0}m_{\tilde{g}}$
($m_{\tilde{g}}$ gaugino mass, like gluino, photino or zino). 

A $\mathcal{W}_{n\psi}$ can be obtained in 
a D-brane model with extra vector-like pairs 
of color-triplets, through the generation 
of a non-perturbative mass term for these, as described in our papers cited above.
But here, we would like to suggest another possibility:
$\mathcal{W}_{n\psi}$ can be {\it directly generated}
from exotic instantons, without the needing of colored-mediators!
In Fig.2, we show an $E2$-instanton directly generating 
the neutron-nweight portal, through mixed disk amplitudes.
In particular: $D^{c}$ comes from excitations
of open strings attached to one $U(1)$-stack (named {\bf a})
and a $U(3)_{c}$-stack (named {\bf c});
$U^{c}$ from $U(1)'$-stack (named {\bf b}) and 
$U(3)_{c}$-stack; $S$ from $U(1)$-stack ({\bf a}) and $U'(1)$-stack ({\bf b}). 
Let us remind that we are considering stacks of D6-branes 
wrapping 3-cycles on $CY_{3}$. 
The E2-brane intersects three times 
the {\bf a}, two times the {\bf b} and 
one time the {\bf c}.  
Effective interactions among ordinary superfields 
$S,U^{c},D^{c}$ and modulini $\alpha^{f=1,2,3},\beta^{g=1,2},\tau^{i=1,2,3}$ 
(living between ordinary D6-branes and $E2$-branes):
\begin{equation}
\label{starting}
{\mathcal{L}}_{E2}\sim Y^{(1)}_{fg}S\alpha^{f} \beta^{g}+Y^{(2)}_{f}\alpha^{f} D^{c}_{i}\tau^{i}+Y^{(3)}_{g}\beta^{g} U^{c}_{i}\tau^{i}
\end{equation}
where $i$ is the color index of $U_{c}(3)$, $f=1,2,3$ labels the number of $\alpha$,
$g=1,2$ the number of $\beta$; $Y^{(1,2,3)}$ are Yukawa matrices
coming from mixed disk correlators.  
Integrating-out modulini
\begin{equation}
\label{starting}
\int d^{3}\tau d^{3}\alpha d^{2}\beta e^{\{Y^{(1)}_{fg}S\alpha^{f} \beta^{g}+Y^{(2)}_{f}\alpha^{f} D^{c}_{i}\tau^{i}+Y^{(3)}_{g}\beta^{g} U^{c}_{i}\tau^{i}\}}
\end{equation} 
we obtain the desired superpotential, with $\mathcal{M}_{0}=M_{S}e^{+S_{E2}}$. 
Such a mechanism can be embedded in D-brane models
like $U_{3}(3)\times U_{L}(2) \times U(1)\times U(1)' \times G$
or  $U_{3}(3)\times Sp_{L}(2) \times U(1)\times U(1)' \times G$,
where $G$ is a generic gauge extension obtained by the global D-brane construction
\footnote{$G$ naturally could be a parallel intersecting D-brane world.
This can intriguing connections with Asymmetric Mirror dark matter phenomenology 
in direct detection \cite{Addazi:2015cua}. }. 
A complete classification of all consistent quivers is beyond the purpose of
this short paper. 
Finally, let us comment that a CPTV mass term for $\psi$ can be introduced in our 
susy model,
as a soft susy/CPT breaking parameter by "diagonal" R-R or NS-NS stringy fluxes
\footnote{The author of this paper retains more probable such a mechanism 
for a CPT breaking with respect to decoherence quantum gravity effects
from virtual miniblack holes, violating quantum mechanics principles. 
See \cite{Addazi:2015gna} for related discussions on these aspects.}.

\subsection{Proton is stable and CPTV in $K_{0}-\bar{K}_{0}$ are smaller than in $n-\bar{n}$}
Now let us comment two important phenomenological aspects of our model.
First, a $\Delta B=1$ superpotential like $U^{c}D^{c}D^{c}S/\mathcal{M}_{0}$
could be dangerous: why not introduce other  $\Delta B,\Delta L=1$ interactions between $S$ 
and SM, immediately destabilizing the proton? 
The answer is because our mechanism has {\it dynamically broken} 
an initial $R$-parity, 
generating one and only one superpotential $U^{c}D^{c}D^{c}S/\mathcal{M}_{0}$.
Other ones are not generated. 
See also our recent papers cited above for discussions on these aspects. 
This one, alone, cannot destabilize the proton.  
Let us note that also other discrete symmetries in the hidden sector can be introduced:
in this case the exotic instanton dynamically breaks also these ones.
In other words, $S$ not interacts with SM at perturbative level
(we have defined it a hidden particle for this motivation). 
However, it can interact non-pertubatively through the $E2$-instanton
considered.

Another  comment regards kaons:
one can construct a diagram generating a kaon-antikaon oscillation
from $\mathcal{W}_{n\psi}=U^{c}D^{c}D^{c}S/\mathcal{M}_{0}$.
In fact, calling $\tilde{s}$ the susy scalar partner of $\psi$,
we can obtain an operator $\tilde{u}^{c}d^{c}d^{c}\tilde{s}/\mathcal{M}_{0}$,
while $\tilde{s}$ can have a susy soft mass $m_{s}^{2}\tilde{s}^{\dagger}\tilde{s}+h.c$.
For our construction, $u,d$ can be up/down-like quarks, but they can have different 
flavors ($c,s,t,b$ quarks).
From these, we can obtain a D-term like diagram for $K_{0}-\bar{K}_{0}$
and in general a plethora of neutral meson oscillations. 
In our model set-up, it seems that CPTV phases propagate also in these channels. 
However, let us note that, these diagrams can be very suppressed 
with respect to $n-\bar{n}$ one:
i) these have an extra one-loop suppression from integration of one squark and one $\tilde{s}$;
ii) $m_{s}$ is a free-parameter that in principle can be putted also up to the Planck scale!
iii) Stringy mixed disk amplitudes are not necessary {\it democratic} with flavor: 
coefficients $Y^{(1,2,3)}$ are Yukawa matrices with flavors.
For these motivations, even if measures in mesons physics remained motivated
in our model, they are expected to be strongly suppressed with respect to neutron ones!

\section{Further implications in $n-\bar{n}$}

In this section, we will comment previous results, 
with possible implications in next generation of experiments 
on $n-\bar{n}$-transitions.

First, let us note that expression (\ref{e1}) implies
\begin{equation}
\label{starting}
R_{CPTV/CPTP}=\frac{\mu_{n\bar{n}}^{CPTV}}{\mu_{n\bar{n}}^{CPTP}}=\frac{m_{\psi}^{2}-\mu_{\psi}^{2}}{-\Lambda_{CPTV}^{2}+m_{\psi}^{2}-\mu_{\psi}^{2}}
\end{equation} 
 If $\Lambda_{CPTV}<<|\mu_{\psi}-m_{\phi}|$, $R_{CPTV/CPTP}\simeq 1$: CPTV
in neutron-antineutron transition are strongly suppressed.

Let us consider a more interesting regime: 
$\Lambda_{CPTV}>>|\mu_{\psi}-m_{\psi}|$.
Usually a neutron-antineutron 
experiment is done in condition of a strongly suppressed magnetic 
field. But in condition of $\Delta m_{n\bar{n}}\neq 0$, 
it is more appropriate to test a $n-\bar{n}$ transition 
with $|\mu_{n} B|\simeq \Delta m_{n\bar{n}}$.
In fact, in this case the transition probability is resonantly enhanced rather 
than suppressed \cite{MagneticAddazi}. 
For example let us suppose that $\Delta m_{n\bar{n}}\simeq  10^{-14}\div 10^{-6}\, \rm eV$:
this case corresponds to a range of external magnetic fields $|B|\simeq 10^{-3}\div 10^{5}\rm Gauss$. 
Clearly, a test with very high magnetic fields up to $1\div 10\, \rm Tesla$ is technologically challenging,
but a test with $0.1\div 1\rm Gauss$ seems simpler to be realized. 
From (\ref{e2}), we can estimate that $\Delta m_{n\bar{n}}\simeq 2\mu_{n\psi}^{2}/\Lambda_{CPTV}$,
that in our instanton-mediated model corresponds to 
$2\Lambda_{QCD}^{6}/\Lambda_{NP}^{5}$,
where $\Lambda_{NP}^{5}=(\mathcal{M}_{0}^{2}m_{\tilde{g}}^{2}\Lambda_{CPTV})\simeq 0.1 \div 10\, \rm TeV$.
This scale is easy to obtain under several reasonable choices of parameters. 
For example, $\mathcal{M}_{0}\simeq 10^{3}\, \rm TeV$, $m_{\tilde{g}}\simeq 1\, \rm TeV$,
$\Lambda_{CPTV}\simeq 0.1\div 10\, \rm MeV$.
In this case, we can marriage our scenario with a low string scale scenario $M_{S}\simeq 1000\, \rm TeV$, with gaugini
reachable at LHC. Otherwise, another possible scenario can be 
$\mathcal{M}_{0}\simeq 10^{3}\, \rm TeV$, $m_{\tilde{g}}\simeq 10^{3}\, \rm TeV$,
$\Lambda_{CPTV}\simeq 0.1\div 10\, \rm eV$. Clearly, in these cases $\psi$ has 
small Lorentz Invariant masses $|m_{\psi}-\mu_{\psi}|<<10^{-1}\div 10^{7}\, \rm eV$. 
Let us note for example that for $\mu_{\psi}\simeq 10^{-19}\div 10^{-9} \Lambda_{CPTV}$,
one can obtain an intriguing situation for $n-\bar{n}$ transitions in presence of external magnetic fields:
$\mu_{n\bar{n}}\simeq 10^{-23\div 25}\, \rm eV$ and 
$\Delta m_{n\bar{n}} \simeq 10^{9 \div 19} \mu_{n\bar{n}}$.
However, a so light $\psi$ seems dangerous: it can lead to meson decays 
like $K^{0}\rightarrow \psi \bar{\psi},\psi\psi$. These decays can be obtain 
with a one loop diagram of squarks. 
However, such a decay has a very suppressed rate for $\Lambda_{CPTV}\simeq 1\,\rm eV \div 1\, \rm MeV$:
$Br(K_{S(L)}\rightarrow \psi\bar{\psi}) \sim \lambda_{\psi}^{2}/m_{K_{S,L}}^{2}(\Lambda_{QCD}/\Lambda_{n\psi})^{2}\mathcal{S}<10^{-31}\div 10^{-17}$,
where $\lambda_{\psi}$ is the mass eigenvalue of $\psi$,
and $\mathcal{S}$ represents other understood suppressions 
in our one-loop diagrams.
The current  limits extracted from existing data are only
$Br(K_{S(L)}\rightarrow invisible)<1.1\times 10^{-4}(6.3 \times 10^{-4})$ \cite{Ginenko}.
These limits are planned to be improved by $2-3$ orders of magnitude.
In particular the planned experimental limits
The planned experimental  limits on $K\rightarrow invisible$ are expected to be at the level of 
$B(K_{S(L)}\rightarrow invisible)<10^{-8}(10^{-6})$ or below,  comparable with other possible meson decays into invisible channels \cite{Ginenko}.
On the other hand, possible double decays $nn\rightarrow \psi\psi$ 
are automatically strongly suppressed: 
$\Gamma_{nn\rightarrow \psi\psi}\sim (\rho_{N}/m_{n}^{2})(\Lambda_{QCD}/\Lambda_{n\psi})^{12}$
with $\rho_{N}\simeq 0.23\, \rm fm^{-3}$,
corresponding to 
$\tau_{nn\rightarrow \psi\psi}\sim 2\times 10^{45}\,\rm yr(\Lambda_{n\psi}/TeV)^{12}$.
A possible nuclei destabilization decays from $n\rightarrow \psi$ is automatically avoided by energy conservation.
A transition $n\rightarrow \bar{n}$ in nuclei is strongly suppressed by the nuclear binding energy $V\simeq 10\div 100\, \rm MeV$. 
However, according to the analysis done in \cite{Remark1}, a case in which $\Lambda_{CPY}>>\mu_{\psi},m_{\psi}$
leads to instabilities in the hidden sector \footnote{Another comment can be that such a scenario just seems to displace the fine-tuning of a CPTV mass
for the neutron to the fine-tuning of a so small CPTV mass for the nweight.
But regardless of our aesthetic prejudices, we have analyzed these cases as a possible
region of the parameter space. }.
However, the analysis done in \cite{Remark1} can be avoid by string-inspired spontaneously or dynamically
mechanisms, involving higher order self-interaction couplings, as specified in the same paper. For example, higher-derivatives 
can relevantly modify the dispersion relation of the nweight in the high energy limit while 
they are completely irrelevant in low energy limits. Such an issue also involves 
deep principles like non-locality. 
To interpret such an effective low energy model in a consistent framework will deserve future investigations. 

In the following part, we will focus on more natural models, compatible with bounds of \cite{Remark1}.

Let us suppose that $m_{\psi}<<\Lambda_{CPTV}\simeq \mu_{\psi}$, in condition of suppressed magnetic fields. This case is free by instabilities, according to the analysis done in \cite{Remark1}.
In this case, $R_{CPTV/CPTP}\simeq 1/2$. 
This also implies a $|\Delta m_{n\bar{n}}|\simeq \mu^{2}_{\psi n}/\Lambda_{CPTV}$.
On the other hand, an insightful way to rewrite these expressions
is in term of $\mu_{n\bar{n}}/|\Delta m_{n\bar{n}}|$:
\begin{equation}
\label{insight}
\frac{\mu_{n\bar{n}}}{|\Delta m_{n\bar{n}}|}=\frac{\mu_{\psi}}{2\Lambda_{CPTV}}
\end{equation}
As a consequence, a situation in which $\Delta m_{n\bar{n}}\simeq \mu_{n\bar{n}}$
corresponds to $\Lambda_{CPTV}\simeq \mu_{\psi}/2$. 
The actual limit on $\mu_{n\bar{n}}$ corresponds to $10^{-23}\, \rm eV$.
 $\mu_{\psi n}=\Lambda_{QCD}^{3}/\Lambda_{n\psi}^{2}$,
where $\Lambda_{n\psi}$ is the New Physics scale in which such a $n-\psi$ vertex is generated, 
As a consequence $\mu_{n\bar{n}}\simeq \Lambda_{QCD}^{6}/(3\Lambda_{n\psi}^{4}\Lambda_{CPTV})$,
under the assumption $m_{\psi}<<\Lambda_{CPTV}\simeq \mu_{\psi}$.
So, $\mu_{n\bar{n}}<10^{-23}\, \rm eV$ corresponds to $\Lambda_{n\psi}^{4}\Lambda_{CPTV}> (100\, \rm TeV)^{5}$.
The next generation of experiments will test $\Lambda_{n\psi}^{4}\Lambda_{CPTV}\simeq (1000\, \rm TeV/3)^{5}$ scale.
A possible interesting scenario can be $\Lambda_{n\psi}\simeq  \Lambda_{CPTV}\simeq \mu_{\psi}/2 \simeq 300\, \rm TeV$.
Such a scenario is easy to obtain in our instanton-mediated model:
$\Lambda_{n\psi}^{4}=[e^{+S_{E2}}M_{S}]^{2}m_{\tilde{g}}^{2}$ so that 
$e^{+S_{E2}}M_{S}\simeq m_{\tilde{g}}\simeq 300\, \rm TeV$ 
as well as $m_{\tilde{g}}\simeq 1\, \rm TeV$ and $e^{+S_{E2}}M_{S}\simeq 10^{5}\, \rm TeV$
can be interesting. In both cases, $M_{S}$ is much smaller than the Planck scale.
Let us  comment that
if $M_{S}\simeq 100\div 1000\, \rm TeV$,
the hierarchy problem of the Higgs mass will be alleviated 
from $m_{H}^{2}/10^{38}\, \rm GeV^{2}\simeq 10^{-34}$ to $m_{H}^{2}/10^{4\div 6}\, TeV^{2}\simeq 10^{-6}\div 10^{-8}$.
As a consequence, a possible future detection of $n-\bar{n}$-transition eventually may suggest 
a future plan for $100\div 1000\, \rm TeV$ proton-proton colliders after LHC. 
On the other hand, because of $e^{S_{E2}}$-factor, $M_{S}=10\, \rm TeV$
remains compatible with our space of parameters: 
LHC will immediately test this scenario in the next run. 
Another possible case can be $\Lambda_{n\psi}\simeq 10^{12\div 13}\, \rm TeV$
while $\Lambda_{CPTV}\simeq 100\div 10^{3}\, \rm GeV$.
In this case, $e^{+S_{E2}}M_{S}\simeq m_{\tilde{g}}\simeq 10^{12\div 13}\, \rm TeV$.

Now, let us discuss another plausible scenario:
$m_{\psi}\simeq \mu_{\psi}$.
In this case, $\mu_{n\bar{n}}\simeq \mu_{n\psi}^{2}\mu_{\psi}/\Lambda_{CPTV}^{2}$.
Region of the parameters are practically the same discussed later, 
replacing $3\Lambda_{CPTV}\rightarrow \Lambda_{CPTV}$.
But in this case, $n-\bar{n}$ is entirely generated by $\Lambda_{CPTV}$.
On the other hand, $|\Delta m_{n\bar{n}}|\simeq 2\mu_{\psi n}^{2}/\Lambda_{CPV}$,
so that $\mu_{n\bar{n}}\simeq |\Delta m_{n\bar{n}}|/2$.
In this case, actual limits on $\Delta m_{n\bar{n}}$ are roughly $10^{-23}\, \rm eV$
\footnote{Let us note that the superpotential generated by exotic instantons (\ref{starting})
has a Yukawa matrix with respect to flavors' indices. This operator can also induce $\Lambda^{0}-\bar{\Lambda}^{0}$ transitions,
violating $B$ as $\Delta B=2$. 
As a consequence, our model inevitably predicts the propagation of CPTV phases 
also in this channel. However, mixed disk amplitudes considered are generically not democratic 
with flavors. As a consequence, CPTV phases' ratio of $\Lambda^{0}-\bar{\Lambda}^{0}$
and $n-\bar{n}$ has an extra parameter depending on the particular homologies
in mixed disk amplitudes considered. This can have relevant implications 
in $\Lambda^{0}-\bar{\Lambda}^{0}$ physics. Possibilities 
to constrains CPTV phases in $\Lambda^{0}-\bar{\Lambda}^{0}$ sector 
were proposed in \cite{Kang:2009xt}: our model can be considered
as a theoretical framework motivating these phenomenological discussions.
However, as mentioned above, one has also to consider 
energy dependences of CPTV phases. Probably, these dependences
can be also neglected in these channels, but these aspects 
will deserve future investigations beyond the purposes of this short paper. 
On the other hand, strong indirect constrains related to this $\Delta B=2$
channel are coming from dinucleon decays $NN\rightarrow KK$:
$100\,TeV$ on the new physics scale, against $100\, GeV$ from electron-positron colliders.
I am grateful to Xianwei Kang for useful informations and remarks in these 
subjects. }.

\section{Conclusions and remarks}

In this paper, we have shown a mechanism 
for a dynamical CPTV, mediated from a hidden 
sector to the neutron (antineutron) one.
In particular, a sterile partner of the neutron 
$\psi$ can be introduced.
We have called such a particle "neutronic weight":
it is a weight for our desired CPTV see-saw.
 An effective neutron-nweight vertex 
transmits informations about the hidden CPTV
to the neutron sector. 
As shown, such a vertex can be induced 
by exotic instantons, without provoking other dangerous ones.
As a consequence, a CPTV mass 
term for the neutron is generated.
We remark that such a mass term is naturally small,
but testable in the next generation of experiments 
in $n-\bar{n}$-transitions! 
On the other hand, such mechanism can be related to 
a primordial CPTV baryogenesis 
through $\psi$-decays $\psi \rightarrow udd,\bar{u}\bar{d}\bar{d}$
or in three quarks collisions $udd\rightarrow \bar{u}\bar{d}\bar{d}$. 
In this case, Sakharov's conditions are not satisfied
and Baryon-violations can be also generated in thermal equilibrium. 
See \cite{CPTVBaryo} for discussions of these aspects in other contests 
and in generic CPTV models.

On the other hand, our model suggests a direct test in 
future colliders. In fact, exotic instantons can be resonantly 
produced in $100-1000\, \rm TeV$ collisions 
like $udd,uds,ubs,..,\rightarrow \psi$. 
In this case, a distinct 'cutoff' in correspondent cross sections,
related to effective lenght scale of the exotic instanton, 
can provide a hint against other gauge models.  

For these motivations, we conclude that 
so mild limits on neutron physics could hide 
unexpected CPTV effects, despite 
of theoretical prejudices:
only experimental observables will have the "last word" about it. 

\begin{acknowledgments} I would like to thank my collaborators Zurab Berezhiani and Massimo Bianchi
 for interesting conversations on these subjects. 
 I also would like to thank Parsa Ghorbani, Sergei Gninenko, Alan Kostelecky, Xianwei Kang and Augusto Sagnotti
 for 
 interesting and useful comments. 
 
My work was supported in part by the MIUR research
grant "Theoretical Astroparticle Physics" PRIN 2012CPPYP7.
\end{acknowledgments}

\end{document}